\documentclass[aps,prb,10pt,twocolumn,groupedaddress,preprintnumbers,amsmath,amssymb,floatfix,longbibliography]{revtex4-2}

\usepackage[english]{babel}
\usepackage[utf8]{inputenc}
\usepackage{newunicodechar}
\newunicodechar{∼}{\sim}

\usepackage{graphicx}
\usepackage{float}
\usepackage{tabularx}
\usepackage{array}
\usepackage{booktabs}
\usepackage{dcolumn}
\usepackage{makecell}
\usepackage{multirow}
\usepackage{bm}
\usepackage{color}
\usepackage[dvipsnames]{xcolor}
\usepackage{braket}
\usepackage{appendix}
\usepackage{upgreek}
\usepackage{chemmacros}
\usepackage{natbib}
\usepackage{soul}
\usepackage{mathrsfs}

\usepackage{hyperref}
\hypersetup{
    colorlinks=true,
    citecolor=blue,
    urlcolor=blue,
    linkcolor=blue
}

\begin{document}

%\preprint{APS/123-QED}  

\title{
Effect of symmetry breaking on altermagnetism in CrSb and Formation of Fragmented Nodal Curves
}   

\author{Arindom Das}
% \affiliation{Condensed Matter Theory and Computational Lab, Department of Physics, IIT Madras, Chennai-600036, India}
% \affiliation{Center for Atomistic Modelling and Materials Design, IIT Madras, Chennai-600036, India}

\author{Arijit Mandal}
% \affiliation{Condensed Matter Theory and Computational Lab, Department of Physics, IIT Madras, Chennai-600036, India}
% \affiliation{Center for Atomistic Modelling and Materials Design, IIT Madras, Chennai-600036, India}

\author{Nayana Devaraj}
% \affiliation{Condensed Matter Theory and Computational Lab, Department of Physics, IIT Madras, Chennai-600036, India}
% \affiliation{Center for Atomistic Modelling and Materials Design, IIT Madras, Chennai-600036, India}

\author{B.R.K. Nanda}
\email{nandab@iitm.ac.in}

\affiliation{Condensed Matter Theory and Computational Lab, Department of Physics, IIT Madras, Chennai-600036, India}
\affiliation{Center for Atomistic Modelling and Materials Design, IIT Madras, Chennai-600036, India}

\begin{abstract}
Phenomena concerning altermagnets have opened up a window for unconventional analysis of the momentum space spin polarization (MSSP) of antiferromagnetic materials. Taking the example of one of the widely investigated altermagnets, CrSb, we explore the underlying mechanisms leading to the formation or breaking of altermagnetism. With the aid of DFT calculation and symmetry analysis, we study the behavior of MSSP in the altermagnetic bands of pristine CrSb, along with a few model structures designed from the pristine one by hypothetical vacancy engineering and interstitial doping. We show that the six-fold rotational symmetry of the pristine CrSb can be reduced to a two-fold rotational symmetry via vacancy and doping engineering. We discover the formation of fragmented nodal curves (FNCs) across the Brillouin zone when in an altermagnetic material when the symmetry is restricted to two-fold rotation. Unlike the typical nodal planes and axes, the location of the FNCs in the momentum space is found to be band-specific. The formation of FNCs is further validated by introducing uniaxial strain to CrSb and by examining the band structure of RbMnPO$_4$, as they both exhibit a two-fold rotational symmetry responsible for altermagnetism. We observe that, unlike the pristine case, these FNCs have the potential to manifest anomalous Hall conductivities (AHC), while the Néel vector orients along both in-plane and out-of-plane directions. This flexibility of the AHC will pave the way for the application of altermagnets in the futuristic quantum devices.

\end{abstract}

\maketitle

\section{Introduction}
Altermagnetism \cite{PhysRevX.12.040501, PhysRevX.12.031042, doi:10.7566/JPSJ.88.123702, bhowal2025}, a third type of magnetic phase, is defined by the splitting of the otherwise degenerate antiferromagnetic sub-bands in the momentum space outside certain nodal planes. The split, often referred to as altermagnetic spin splitting (AMSS), which resembles the time reversal symmetry breaking as in ferromagnets, is equal and opposite on either side of the nodal plane and axis to ensure zero magnetization when integrated over the whole Brillouin zone (BZ) as in an antiferromagnet (AFM). Being purely a non-relativistic quantum phenomenon, with unconventional momentum space spin polarization (MSSP) that deviates from both classical ferromagnets and AFMs, altermagnets have the potential to open up non-trivial routes for quantum transport in the field of spintronics \cite{adma.202505779, PhysRevX.12.011028, adfm.202409327}.

Among the most investigated altermagnets, prominent ones include the NiAs prototype transition-metal-based alloys such as  MnTe\cite{PhysRevLett.130.036702, PhysRevLett.132.036702, PhysRevMaterials.8.104407, Nanoscale_imaging}, CrSb\cite{Reimers2024, priya_CrSb, PhysRevLett.133.206401, Yang2025}, and NiS\cite{sm63-1dcx}. These high-symmetry compounds, which stabilize in the hexagonal P6$_{3}$/mmc phase, are either correlated insulators or simple metals and exhibit a large AMSS of the order of 1 eV. Furthermore, doping and defects can lead to quasi-altermagnetism with uncompensated magnetization \cite{devaraj2025unlockingdopingeffectsaltermagnetism} as well as finite anomalous Hall conductivity (AHC) \cite{zero_field_AHE, adfm.202508282}. Therefore, these compounds provide an appropriate platform to study the underlying mechanisms responsible for forming the altermagnetism and tailoring the latter by designing several symmetry-breaking model structures. We pursued this objective in this work by taking CrSb as the example system.

CrSb exhibits many intriguing properties other than altermagnetism. In the bulk phase, it has a large magnetocrystalline anisotropy energy (MAE) \cite{PhysRevB.102.224426}, which is desirable for storing information. The MAE of CrSb is reduced in the thin films \cite{https://doi.org/10.1002/adma.202508977}, which helps in manipulating the Néel vector and, thereby, AHC through magnetic proximity. The reorientation of the Néel vector can also be achieved using strain, electric field, doping, or voltage \cite{zhou2024crystaldesignaltermagnetism, fischer2026engineering, giri2026straintopologicalselectoraltermagnetic}, and it is an emerging topic as this is identified as one of the routes to induce tunable anomalous Hall \cite{ahc1, ahc2} and Nernst conductivities \cite{Zhou2025, Yu2025}. Additionally, it exhibits moderate energy barriers between its altermagnetic, ferromagnetic, and ferroelectric phases \cite{adfm.202525978} making it more flexible for multiferroic spintronics.

Furthermore, a recent study shows that breaking the crystal and magnetic symmetry using doping or strain can be employed to tune the unconventional spin Hall effect, allowing for possible applications in various spintronics \cite{Jeong}. In addition, the coexistence of topological order and altermagnetism is being actively investigated in CrSb. From the ARPES data, it is shown that the large AMSS induces altermagnetic Weyl nodes and topological Fermi arcs, which suggests robust electronic topology in this compound \cite{Li2025}. Reports suggest that by utilizing laser-induced ultrafast demagnetization on CrSb, promising applications in ultrafast spintronics can be realized \cite{15ct-lzds}. 

In this work, we have carried out first-principles electronic structure calculations on pristine and strained CrSb, and derived hypothetical model structures (MSs) where vacancies and interstitial doping are done at the non-magnetic sites to alter the chemical bonding and thereby to break the symmetry of charge and spin distributions. This allows us to study the evolution of MSSP of the bands with change in the symmetry. 

The MSs allowed us to explore how variations in the doping concentration of non-magnetic atoms alter the local spin moment (LSM) of the system, subsequently impacting the AMSS in reciprocal space. We observed that systems with improper rotation, such as S$_{6z}$, create similar AMSS as those formed by proper rotation C$_{6z}$. This phenomenon has been discussed and mathematically proven in the non-relativistic limit (NRL) for pristine CrSb. An interesting fundamental phenomenon occurs when the symmetry is lowered from $C_{6z}/S_{6z}$ to C$_{2z}$. Most remarkably, in this process, the three diagonal nodal planes vanish, and fragmented nodal curves (FNCs) emerge. Along the FNC, the spin-opposite sublattice pair bands are degenerate; elsewhere, they are non-degenerate. For each pair of bands, we observe different FNCs. The lowering in symmetry also allows tuning the Néel vector orientations and realizing a finite AHC both in-plane and out-of-plane. In practice, this can be achieved by uniaxial straining of CrSb.

\begin{figure*}[hbt!]
    \centering  
    \includegraphics[scale = 0.575]{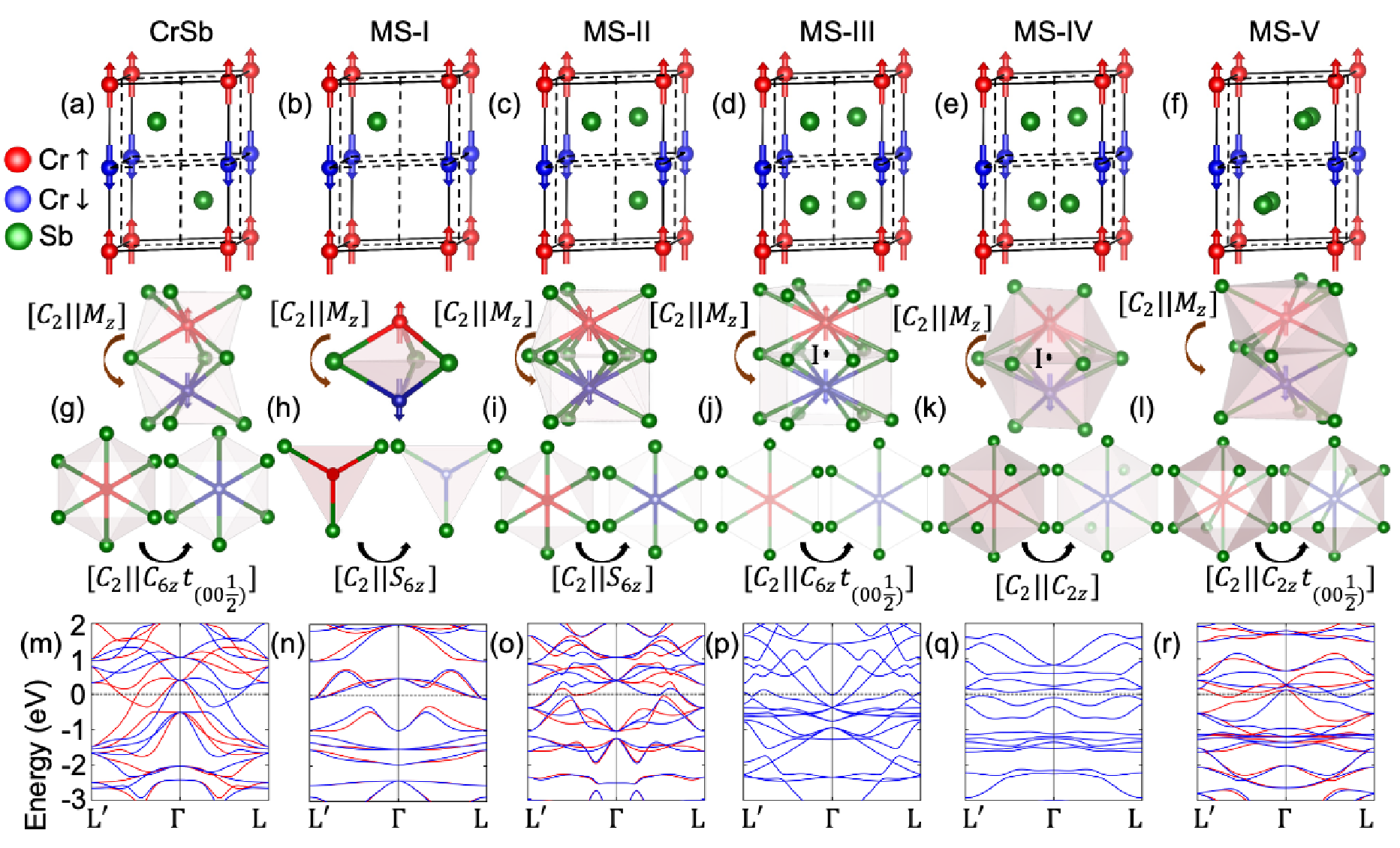}
    \caption{The effect of vacancy engineering and interstitial doping on altermagnetism in CrSb. The top row depicts the crystal structure of (a) pristine CrSb, (b) vacancy-induced configuration Cr$_2$Sb (MS-I), and (c) various doping configurations, such as (d) Cr$_2$Sb$_3$ (MS-II) and (e-g) CrSb$_2$ (MS-III, MS-IV, and MS-V, respectively). The second row schematically illustrates symmetry operations of the spin space group required to define an altermagnet and its nodal planes. The upper subrow indicates the presence of M$_z$ symmetry between two opposite spin sublattices in all configurations, leading to a basal nodal plane, while the lower subrow represents the improper rotation symmetries that define altermagnetism and corresponding diagonal nodal planes (see Fig.~\ref{nodal_planes} (a)). (g) Pristine CrSb has a six-fold screw, and (h, i) MS-I, MS-II have a six-fold rotoinversion symmetry, which creates altermagnetism with three diagonal nodal planes. For (j) MS-III and (k) MS-IV, although there exist a six-fold and two-fold screw, the presence of an inversion centre (I, represented by a black dot) between two opposite spin sublattices forces them to be antiferromagnet. (l) Most intriguingly, MS-V has a two-fold screw, which manifests a non-trivial nodal curve, which we call the fragmented nodal curve. The bottom row represents the altermagnetic (m-o, r) and antiferromagnetic (p, q) band structure of the respective crystal structure of the same column along the high symmetric k-path L$’$-$\Gamma$-L.}
    \label{doping}
\end{figure*}

\section{Structural and Computational Details}

CrSb belongs to the space group P6$_3$/mmc and displays a structural similarity to hexagonal NiAs, as shown in Fig.~\ref{doping} (a). It crystallizes with a two formula unit cell (Cr$_2$Sb$_2$). Its magnetic ordering follows the magnetic space group (MSG) of P6$'_3$/m$'$m$'$c, with its Néel vector oriented along the easy axis, i.e., $z$-axis. In the NRL, the spin space group (SSG) \cite{LITVIN1974538, Litvin:a14103, PhysRevX.14.031037, PhysRevX.14.031038, PhysRevX.14.031039, 10.1098/rspa.1966.0211} notation will be P$^{-1}$6$_{3}$/$^{-1}$m$^{1}$m$^{1}$c$^{\infty m}$1 according to \cite{SSG_nature}, which have been followed throughout this article. Cr and Sb occupy the Wyckoff positions 2a and 2c with the $\bar{3}$m and $\bar{6}$m2 site symmetry \cite{Bilbao}, respectively. The opposite spin sublattices form face-shared octahedra as shown in Fig.~\ref{doping} (g) with a Cr-Sb bond length of 2.74 \AA \, and Sb-Cr-Sb(Sb$'$) angles of 97.36$^{\circ}$(82.64$^{\circ}$). In this hexagonal crystal structure, eight interstitial voids are present. Notably, the centers of two of these voids are occupied by the Sb atom.

\begin{figure}[htbp]
    \centering
    \includegraphics[width=1.0\linewidth]{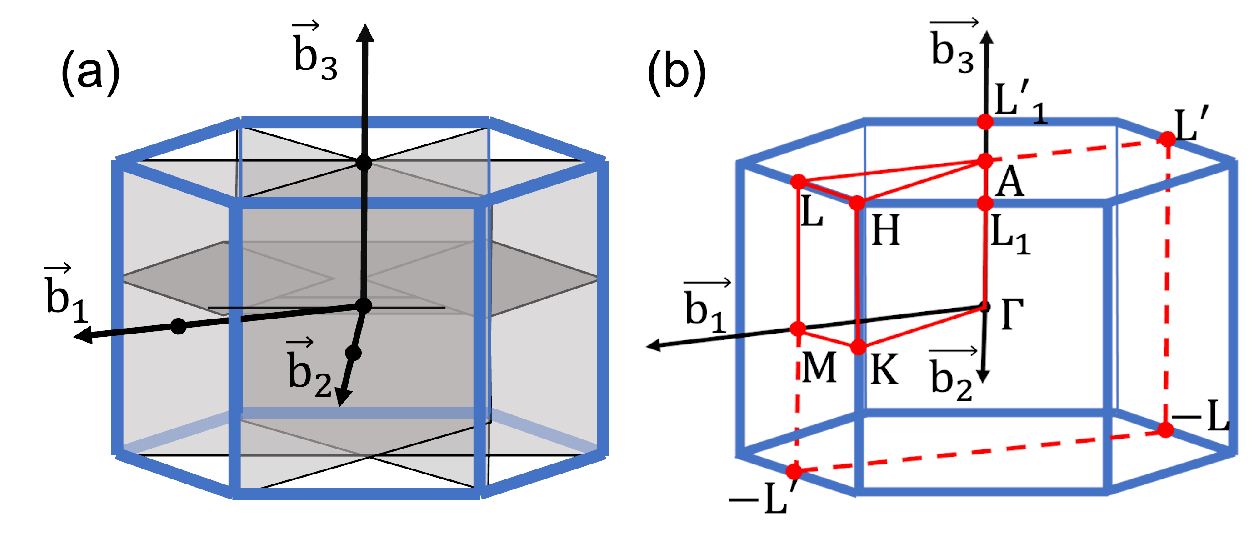}
    \caption{Schematic illustration of various altermagnetic nodal planes present in pristine CrSb, MS-I, and MS-II. (a) represents one basal nodal plane and three diagonal nodal planes, and (b) represents the high symmetry k-points and k-path in this Brillouin zone.}
    \label{nodal_planes}
\end{figure}

The two spin-opposite sublattices are connected via [C$_2$$||$M$_z$] and [C$_2$$||$C$_{6z}$t$_{(0\,0\,1/2)}$] symmetry operations Fig.~\ref{doping} (g) giving rise to the 4 nodal planes in the BZ passing through $\Gamma$ \cite{sm63-1dcx}. One of them is the basal $k_x-k_y$ plane, and the other three nodal planes are perpendicular to it, making an angle of $\pi/3$ with each other. The algebraic equations of these planes are (i) $k_x-k_y = 0$, (ii) $2k_x + k_y = 0$, (iii) $k_x +2k_y = 0$, and (iv) $k_z = 0$  as shown in Fig.~\ref{nodal_planes} (a), written in terms of the hexagonal coordinate system. In the symmetry operation notation [C$_2$$||$C$_{6z}$t$_{(0\,0\,1/2)}$], the operation on the left (right) side of ``$||$'' operates on spin space (real space), i.e.,  C$_2$ is two fold rotation in spin space, while C$_{6z}$t$_{(0\,0\,1/2)}$ is six fold rotation about $z$-axis followed by $(0\,0\,1/2)$ translation and M$_z$ is a mirror operation with respect to $xy$ plane in the real space. 

To investigate the magnetic and electronic structure of the system, Density Functional Theory (DFT) was employed utilizing the plane-wave based projector augmented wave (PAW) \cite{Blochl1994-wp, PhysRevB.59.1758} method as implemented within the Vienna ab-initio simulation package (VASP) \cite{PhysRevB.54.11169}. The Perdew–Burke–Ernzerhof generalized gradient approximation (GGA) \cite{ernzerhof1998generalized} was employed for the exchange-correlation functional. A $\Gamma$-centered k-mesh of $12 \times 12 \times 6$ was utilized for the integration of Brillouin zones (BZ). The kinetic energy cutoff for the plane wave basis set was determined to be 500 eV. For the calculation of RbMnPO$_4$, a dense $\Gamma$-centered k-mesh grid of $6 \times 9 \times 6$ was employed with an energy cutoff value of 500 eV. The strong correlation effects for RbMnPO$_4$ were accounted for using the DFT + U method with the effective Hubbard parameter $U_{\text{eff}} = U - J = 3$ eV in the rotationally invariant functional framework as provided by Dudarev \cite{dudarev}.

To calculate the AHC, first-principles calculations were performed within the framework of Quantum ESPRESSO \cite{giannozzi2009quantum} using norm-conserving pseudopotentials \cite{hamann2013optimized} by incorporating spin orbit coupling.
Maximally localized Wannier functions, obtained using the Wannier90 code, \cite{mostofi2014updated}, were employed to construct the tight-binding Hamiltonian. The anomalous Hall conductivity was then obtained by using the WannierBerri code~\cite{tsirkin2021high}.

\section{Model structures and associated symmetries}

In order to understand the effect of vacancy and interstitial doping on altermagnetism, we have created five MSs out of the pristine crystal structure by adding and removing non-magnetic atoms. The last three, from MS-III to MS-V, are volume relaxed to prevent the Sb atoms from overlapping. In the following paragraphs, we have discussed the crystal structure in details of each MS and summarized the lattice parameters and their SSG in Table~\ref{MS-structre}. 

\textit{Model structure - I (Cr$_2$Sb$_1$):} In this MS, one of the Sb atoms is removed, as shown in Fig.~\ref{doping} (b), to create a crystal system with formula unit Cr$_2$Sb$_1$. The new space group is P$\bar{6}$m2, and Cr and Sb have Wyckoff positions 2g and 1f, respectively. Here, we have 3m. and $\bar{6}m2$ site symmetry for Cr and Sb, respectively. The SSG notation for the system is P$^{-1}$$\bar{6}$$^{1}$m$^{-1}$2$^{\infty m}$1. The two magnetic sublattices are connected via [C$_2$$||$M$_z$] and [C$_2$$||$S$_{6z}$] symmetry as shown in Fig.~\ref{doping} (n). S$_{6z}$ is 6-fold roto-inversion, which is a 6-fold rotation followed by an inversion. %Therefore, like the pristine system, here also we expect altermagnetism with BZ having the same 4 nodal planes (Fig.\ref{nodal_planes}-(a)) as for the pristine compound.

\textit{Model structure - II (Cr$_2$Sb$_3$):} In these hypothetical structures we have occupied one more void with an additional Sb atom. The separation distance between two nearest neighbors is $\sim$ 1.37 \AA\, and between two second neighbors, it is $\sim$ 2.61 \AA. The latter looks more feasible with an additional Sb atom, and there are three possible identical arrangements to create the compound Cr$_2$Sb$_3$. One of them is shown in Fig.~\ref{doping} (c), which forms this MS. The primitive cell of Cr$_2$Sb$_3$ falls under space group P$\bar{6}$m2 and SSG P$^{-1}$$\bar{6}$$^{1}$m$^{-1}$2$^{\infty m}$1, having Cr occupying 2g and three Sb occupying 1c, 1d, and 1f Wyckoff positions. Like Cr$_2$Sb$_1$ system, the opposite spin the sublattices for Cr$_2$Sb$_3$ shown in the Fig.~\ref{doping} (o) are also connected via [C$_2$$||$M$_z$] and [C$_2$$||$S$_{6z}$] symmetry. 

\textit{Model structure - III and IV (Cr$_2$Sb$_4$):} By placing one more Sb in a second neighbor hexagonal center, one can form the compound Cr$_2$Sb$_4$. There are two such possible arrangements, MS-III and MS-IV, as shown in Fig.~\ref{doping} (d) and (e), respectively. MS-III is associated with the space group P6/mmm and SSG  P$^{1}$6/$^{1}$m$^{1}$m$^{1}$m$^{-1}$(0 0 1/2)$^{\infty m}$1. In this model, Cr and Sb occupy 1a and 2d Wyckoff positions showing 6/mmm and $\bar{6}m2$ site symmetry, respectively. The spin opposite sublattices can be connected via the inversion center shown using a black dot in Fig.~\ref{doping} (j). Based on the standard symmetry, it should not exhibit altermagnetism.

The primitive cell of MS-IV follows the space group Cmmm and SSG C$^{1}$m$^{1}$m$^{-1}$m$^{\infty m}$1. The Cr occupies the 4k, and Sb occupies the 4h and 4i Wyckoff positions. Here, the 4k site has mm2 site symmetry, while the 4h and 4i sites have 2mm and m2m site symmetry. Similar to the previous case, the two opposite spin sublattices for this system can also be connected via an inversion center (I) identified in Fig.~\ref{doping} (k) by a black dot, turning this into a pure antiferromagnetic system.

\textit{Model structure - V (Cr$_2$Sb$_4$): } In this model structure, we have maintained the formula unit Cr$_2$Sb$_4$ and placed the Sb atoms in two nearest neighbor hexagonal center positions, as shown in Fig.~\ref{doping} (f). The structure is associated with space group P2$_1$/m and SSG P$^{-1}$2$_1$/$^{-1}$m$^{\infty m}$1. The Cr and Sb here occupy the 2a and 2e Wyckoff positions, maintaining $\bar{1}$ and m site symmetry. The spin opposite sublattices for this case are not connected via any inversion or translation symmetry. From Fig.~\ref{doping} (q) it is clear that the opposite spin sublattices are related via [C$_2$$||$M$_z$] and [C$_2$$||$C$_{2z}$t$_{(0\,0\,1/2)}$] symmetry creating the distinct AMSS type splitting in the BZ. 

\begin{table}[htbp]
\centering
\caption{Magnetic space group (MSG), lattice parameters, and corresponding spin space group (SSG) for pristine CrSb and the five magnetic structures (MS) with the Néel vector along the $z$ axis.}
\label{MS-structre}

\renewcommand{\arraystretch}{1.15}
\setlength{\tabcolsep}{4pt}

\begin{tabular}{l c c l}
\toprule
\toprule
Structure & $a$ (\AA) & $c$ (\AA) & SSG \\
\midrule
CrSb / Cr$_2$Sb$_2$ & 4.11 & 5.45 & $P^{-1}6_{3}/^{-1}m^{1}m^{1}c^{\infty m}1$ \\
MS-I   & 4.11 & 5.45 & $P^{-1}\bar{6}^{1}m^{-1}2^{\infty m}1$ \\
MS-II  & 4.11 & 5.45 & $P^{-1}\bar{6}^{1}m^{-1}2^{\infty m}1$ \\
MS-III & 6.18 & 7.10 & $P^{1}6/^{1}m^{1}m^{1}m^{-1}(0 0 1/2)^{\infty m}1$ \\
MS-IV  & 6.18 & 7.10 & $C^{1}m^{1}m^{-1}m^{\infty m}1$ \\
MS-V   & 6.18 & 7.10 & $P^{-1}2_1/^{-1}m^{\infty m}1$ \\
\bottomrule
\bottomrule
\end{tabular}
\end{table}

\section{Electronic Structure}
\begin{figure}[htbp]
    \centering
    \includegraphics[width=1.0\linewidth]{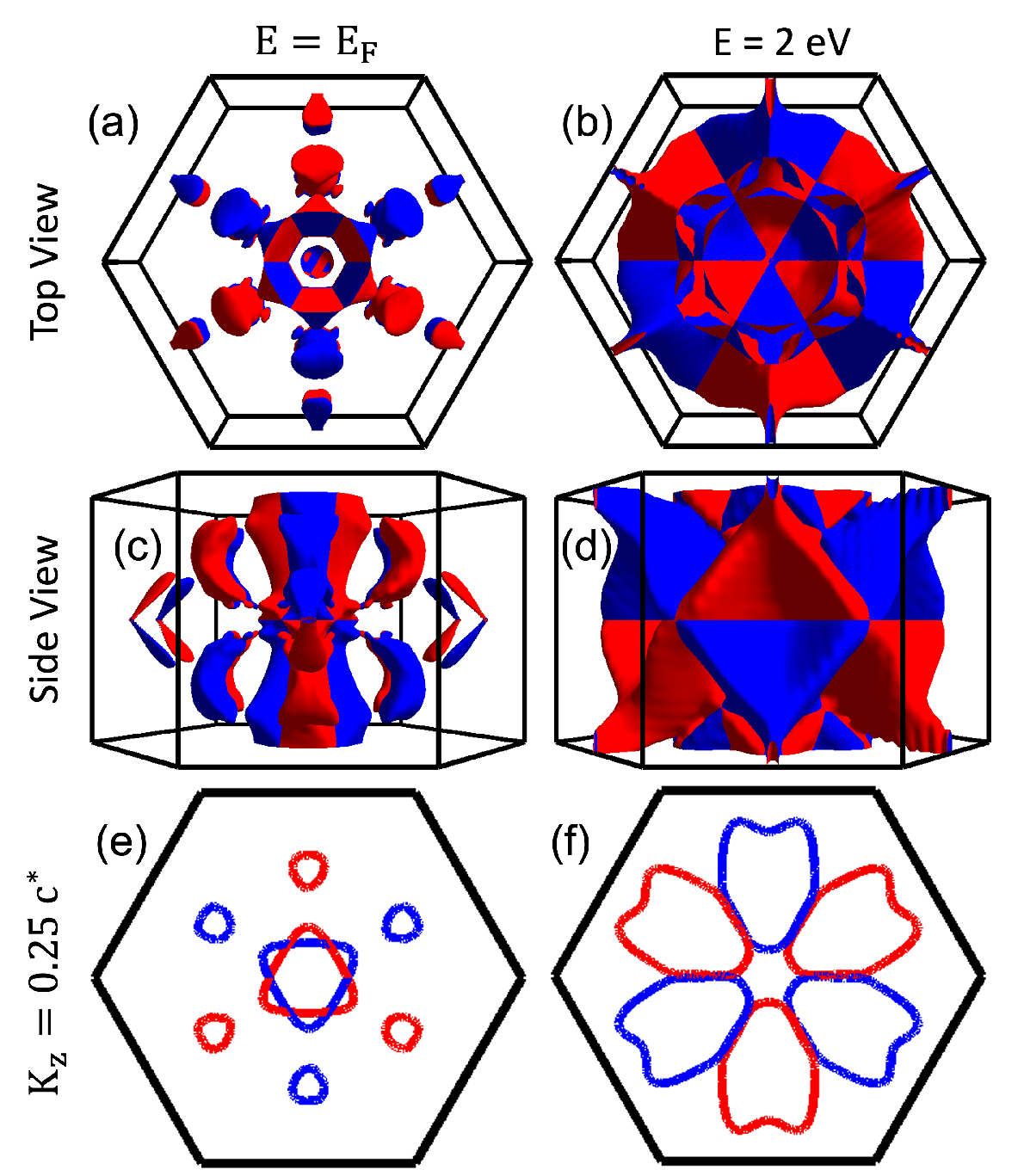}
    \caption{Constant energy surface of pristine CrSb at different energies. The top (a, b) and middle (c, d) panels represent the top and side view of the constant energy surface for a given energy, while the bottom panel (e, f) represents its k$_x$-k$_y$ cross-sectional view at k$_z = 0.25 \, c^{*}$. Each column is dedicated to one particular energy. The left and right columns represent the constant energy surface at 0 and 2 eV, respectively.}
    \label{pristine_fermi_surface}
\end{figure}
The evolution of the antiferromagnetic band structure with change in the symmetries as devised in the MS-I to MS-V is the focus of this section. To start with, we briefly summarize the electronic structure of pristine CrSb, which has been discussed in great detail in the past \cite{Yang2025, Reimers2024}. As shown in Fig.~\ref{doping} (m), the ground state of CrSb exhibits altermagnetic band structure with equal and opposite AMSS on either side of one of the diagonal nodal planes. The constant energy surfaces plotted in Fig.~\ref{pristine_fermi_surface} from different perspectives demonstrate the presence of the four nodal planes emerging from the mirror (M$_z$) and roto-translation (C$_{6z}$t$_{001/2}$). Despite being an alloy, due to the difference in the electronegativity between Sb and Cr, the former (5s$^2$5p$^3$) acts like an acceptor and the latter (4s$^{1}$3d$^{5}$) acts like a donor. The LSM at Cr site is calculated to be 2.73 $\mu_B$, which suggests a high spin configuration with a tentative Cr$^{3+}$ configuration. From the bonding perspective, the origin of altermagnetic band dispersion and strong AMSS  is primarily attributed to the deterministic role of second neighbor interaction among the Sb-$p$ orbitals \cite{sm63-1dcx}. 

\textit{Model structure - I:} From the electronic structure perspective in MS-I, removal of one Sb in Cr$_2$Sb$_2$, more number of electrons are available at the Cr site and at the same time the covalent hybridization is weakened to narrow the bandwidth as can be seen from Fig.~\ref{doping} (n). It leads to the enhancement in the localization which in turn favors the high spin state. Therefore, we observe a higher LSM of $\sim$3.5$\mu_B$. The magnetic ordering of this and the other MS is maintained with the A-type configuration to study the altermagnetic phenomena.

The band structure of MS-I exhibits the AMSS characteristics, and the system has the same four nodal planes due to the presence of S$_{6z}$ roto-inversion and M$_z$ mirror symmetries. However, the AMSS strength is almost one order smaller compared to that of the pristine compound, and this can be attributed to the breakdown of the Sb-Sb second neighbor interaction.

\begin{figure}
    \centering
    \includegraphics[width=1.0\linewidth]{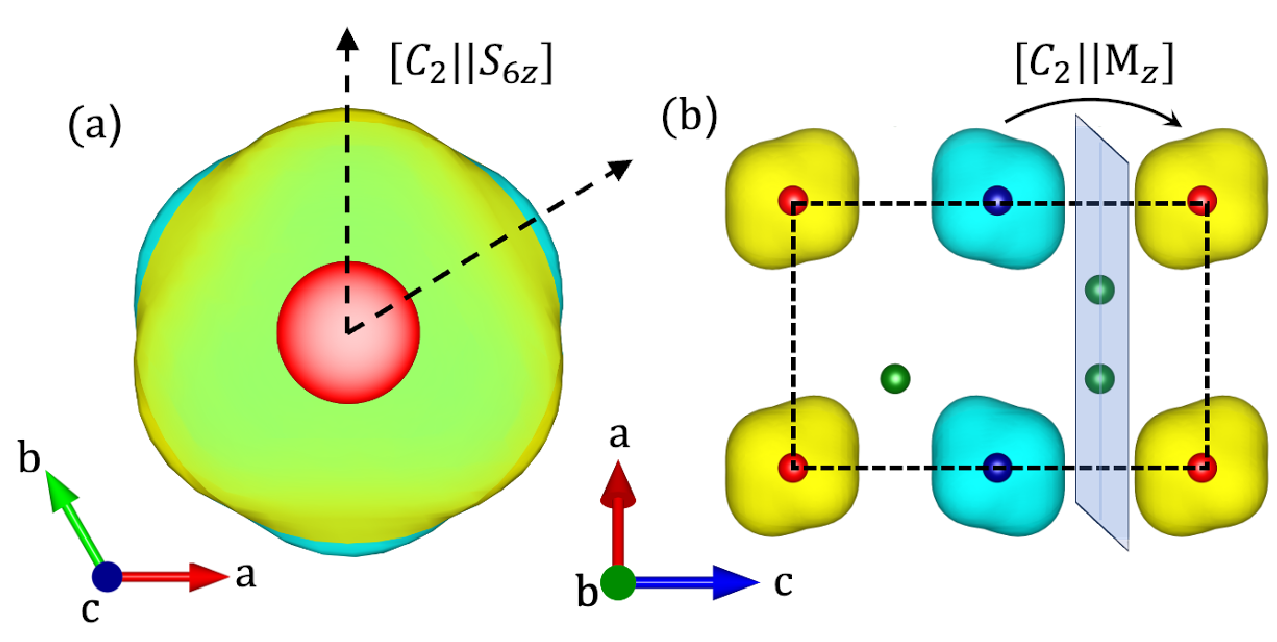}
    \caption{Spin density contours of Cr atoms and their symmetries in MS-II. (a) Top view (from $z$-axis) of spin densities is shown (isosurface = 0.04), which indicates a $\pi/6$ rotation between two opposite spins. (b) Side view (from $y$ axis)  of spin densities is shown (isosurface = 0.008), indicating the presence of M$_z$ symmetry between the opposite spin contours. The yellow and cyan colors represent the up and down spin density contours, respectively.
    }
    \label{spin_density}
\end{figure}

\textit{Model structure - II:} Structure wise, this system can be considered as the pristine $Cr_2Sb_2$ with an additional Sb occupying the center of one of the 6 remaining hexagonal pockets. This adds more valence electrons and hence more bands to the system as can be seen from Fig.~\ref{doping} (o). Compared to the pristine bands, in this MS we also observe narrower bands. It resembles the system where doping and substitutions often create perturbation in the local potential leading to localization of the states. With a smaller number of electrons at the Cr site, the LSM decreases to $\sim$1$\mu_B$. Following the S$_{6z}$ roto-inversion and M$_z$ mirror symmetries, the band structure MS-II has altermagnetic characteristics with reduced AMSS as in the case of MS-I.  

As discussed above, in MS-I and MS-II, the spin opposite sublattices are connected by [C$_2||$S$_{6z}]$ symmetry while in the pristine system they are connected by [C$_2||$C$_{6z}]$ and in both the cases we have four nodal planes which implies that proper and improper rotation connecting the sublattices creates similar effect in reciprocal space while determining the altermagnetic characteristics. The reason for it is as explained below.

In (NRL), time-reversal symmetry is operated on the momentum space while applying identity on spin and this dual operation is denoted as [E$||$T$_{R,L}]$ \cite{PhysRevX.12.031042}. The effect is identical to that of an inversion operator I, and it results in $\epsilon(\vec{k}, \sigma) = \epsilon(-\vec{k}, \sigma)$. Therefore, in the non-relativistic limit, the operations C$_{6z}$ and S$_{6z}$ behave identically as can be seen below from Eq.~\ref{Eq.1} and \ref{Eq.2}. If we define, $\vec{k} = k_x \hat{b}_1+ k_y \hat{b}_2 + k_z \hat{b}_3$, then, 
\begin{equation}
\begin{aligned}
    \psi_{(k_x, k_y, k_z, \sigma)} &\overset{[C_2||C_{6z}]}{\rightarrow} \psi_{(k_x-k_y, k_x, k_z, - \sigma)}, \\
    \psi_{(k_x-k_y, k_x, k_z, - \sigma)} &\overset{[E||T_{R,L}]}{\rightarrow} \psi_{(-k_x+k_y, -k_x, -k_z, -\sigma)},
\label{Eq.1}
\end{aligned}
\end{equation}
and, as S$_{6z} = $ IC$_{6z}$,  
\begin{equation}
    \psi_{(k_x, k_y, k_z, \sigma)}\overset{[C_2||S_{6z}]}{\rightarrow} \psi_{(-k_x+k_y, -k_x, -k_z, - \sigma)}.
\label{Eq.2}
\end{equation}
Therefore, we observe degenerate eigenvalues: 
\begin{equation}
\begin{aligned}
    \epsilon(k_x, k_y, k_z, \sigma) &= \epsilon(k_x-k_y, k_x, k_z, - \sigma) \\
    &= \epsilon(-k_x+k_y, -k_x, -k_z, -\sigma)
\label{Eq.3}
\end{aligned}
\end{equation}
 
Consequently, the four nodal planes observed for the system with C$_{6z}$t$_{(00\frac{1}{2})}$ (Fig.~\ref{nodal_planes}) are also observed for systems with S$_{6z}$ symmetry. These systems generate comparable spin density in reciprocal space as illustrated in Fig.~\ref{spin_density} (a). In Fig.~\ref{spin_density} (b), the presence of a mirror plane with the corresponding spin density is demonstrated from the $y$-axis.

\textit{Model structure - (III - IV):} With formula unit Cr$_2$Sb$_4$, the MS-III and IV are Sb rich systems and behave like ideal alloys where the electron transfer is weak and only electron sharing occurs through hybridization. The d$^4$ electronic configuration stabilizes in a high-spin state to generate an LSM of $\approx$ 4$\mu_B$ at the Cr site. The presence of an inversion center, as shown by the black dots for the corresponding structure in the middle panel of Fig.~\ref{doping} (j) and (k), ensures no altermagnetic characteristics in the band structure (Fig.~\ref{doping} (p) and (q)).

It is trivial to see the spin-opposite sublattice band degeneracy in the presence of an inversion center as,
\begin{equation}
\begin{aligned}
    \psi_{(k_x, k_y, k_z, \sigma)} &\overset{[C_2||I]}{\rightarrow} \psi_{(-k_x, -k_y, -k_z, - \sigma)},
\label{Eq.4}
\end{aligned}
\end{equation}
and in the NRL,
\begin{equation}
\begin{aligned}
\psi_{(-k_x, -k_y, -k_z, - \sigma)} 
&\overset{[E||T_{R,L}]}{\rightarrow} \psi_{(k_x, k_y, k_z, -\sigma)}.
\label{Eq.5}
\end{aligned}
\end{equation}
Therefore,  
\begin{equation}
    \epsilon(k_x, k_y, k_z, \sigma) =
    \epsilon(-k_x, -k_y, -k_z, -\sigma).
\label{Eq.6}
\end{equation}

\textit{Model structure - V:} The MS-V has the same stoichiometry as that of MS-III and IV, and hence it also behaves like an ideal alloy and generates a LSM of $\approx$ 4$\mu_B$ at the Cr site. However, due to the lack of inversion center, and the presence of a mirror symmetry M$_z$, in this case the altermagnetic band structure is observed as can be seen from Fig.~\ref{doping} (r). Furthermore, due to the absence of C$_{6z}$t$_{(00\frac{1}{2})}$ or S$_{6z}$, only the basal nodal plane survives, and on further analysis, we observe the presence of a nodal axis along $\hat{z}$ and passing through $\Gamma$. The origin of formation of a single nodal plane and nodal axis is further discussed in section~\ref{section-FNC}. 

\section{Fragmented nodal curves in compounds with lower symmetry}
\label{section-FNC}
The three nodal planes emerging from the (C$_{6z}$)/(S$_{6z}$) symmetry intersect with each other to create a nodal line along $k_{z}$-axis passing through $\Gamma$. Such a nodal line is seen in the pristine system as well as in MS-I and II. This is conventional and holds the properties of the three nodal planes it is associated with. However, the constant energy surfaces plotted for MS-V demonstrate the existence of random FNCs all across the BZ, which has not been reported earlier. This intriguing phenomenon, which may eventually lead to the practical application of AM materials, can be understood from the following symmetry analysis. The MS-V has the symmetries [C$_2 || $M$_{2z}]$ and [C$_2 ||$C$_{2z}]$, which transforms the wavefunction as:

\begin{equation}
     \psi_{(k_x, k_y, k_z, \sigma)} \overset{[C_2 || M_{2z}]}{\rightarrow} \psi_{(k_x, k_y, -k_z, -\sigma)},
\label{Eq.7}
\end{equation}

\begin{equation}
     \psi_{(k_x, k_y, k_z, \sigma)} \overset{[C_2 || C_{2z}]}{\rightarrow} \psi_{(-k_x, -k_y, k_z, -\sigma)}.
\label{Eq.8}
\end{equation}

However, Eqs.~\ref{Eq.7} and ~\ref{Eq.8} are connected by the NRL led dual operation $[E||T_{R,L}]$, i.e., 
\begin{equation}
    \psi_{(k_x, k_y, -k_z, -\sigma)} \overset{[E || T_{R,L}]}{\rightarrow} \psi_{(-k_x, -k_y, k_z, -\sigma)}, 
\label{Eq.9}
\end{equation}
Therefore, Eqs. (6) - (8), establish the following relation for the eigenvalues,
\begin{equation}
\begin{aligned}
    \epsilon(k_x, k_y, k_z, \sigma) &= \epsilon(-k_x, -k_y, k_z, -\sigma) \\
    &= \epsilon(k_x, k_y, -k_z, -\sigma).
\label{Eq.10}
\end{aligned}
\end{equation}

\begin{figure}
    \centering
    \includegraphics[width=1.0\linewidth]{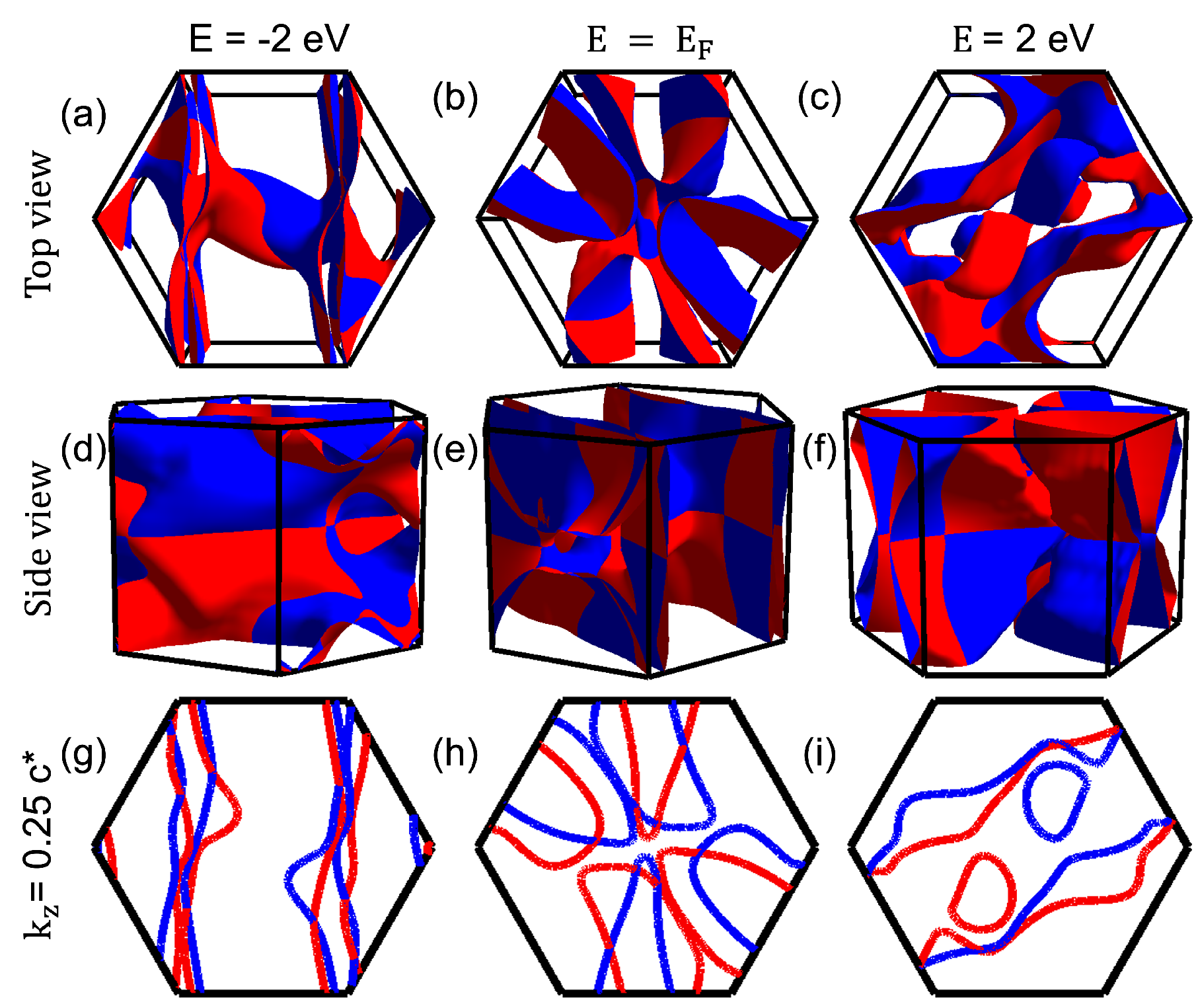}
    \caption{Formation of fragmented nodal curves in MS-V. The top (a-c), middle (d-f), and bottom (g-i) panels represent the top view, side view, and a $k_x-k_y$ cross-sectional view at $k_z=0.25c*$ of the constant energy surface, respectively, for a given energy as explained in Fig. \ref{pristine_fermi_surface}. The left, middle, and right columns represent the constant energy surface at -2, 0, and 2 eV, respectively. The constant energy surfaces clearly indicate the absence of diagonal nodal planes and the formation of FNCs in MS-V.
    }
    \label{3d_fermi_surface}
\end{figure}

The above equation connects the two spin-opposite sublattice bands through a rotation of $\pi$ on a k$_x$-k$_y$ plane for any given k$_z$ (see Fig.~\ref{3d_fermi_surface} (upper panel)) and along k$_z$ for any given k$_x$-k$_y$ (see Fig.~\ref{3d_fermi_surface} (middle panel)). The former implies that two spin-opposite bands (pair bands) have to intersect each other at a given point (nodal point) in the k$_x$-k$_y$ plane. This is demonstrated in the lower panel of Fig.~\ref{3d_fermi_surface} for k$_z = 0.25c^*$. As we vary the k$_z$ continuously, these nodal points form an FNC on the constant energy surface. As the energy varies, the shape of the nodal curve also changes. Furthermore, for each pair of bands, the shape of the nodal curve is different in general.

\subsection{Tailoring FNCs in Pristine CrSb through uniaxial strain}

\begin{figure}
    \centering
    \includegraphics[width=1\linewidth]{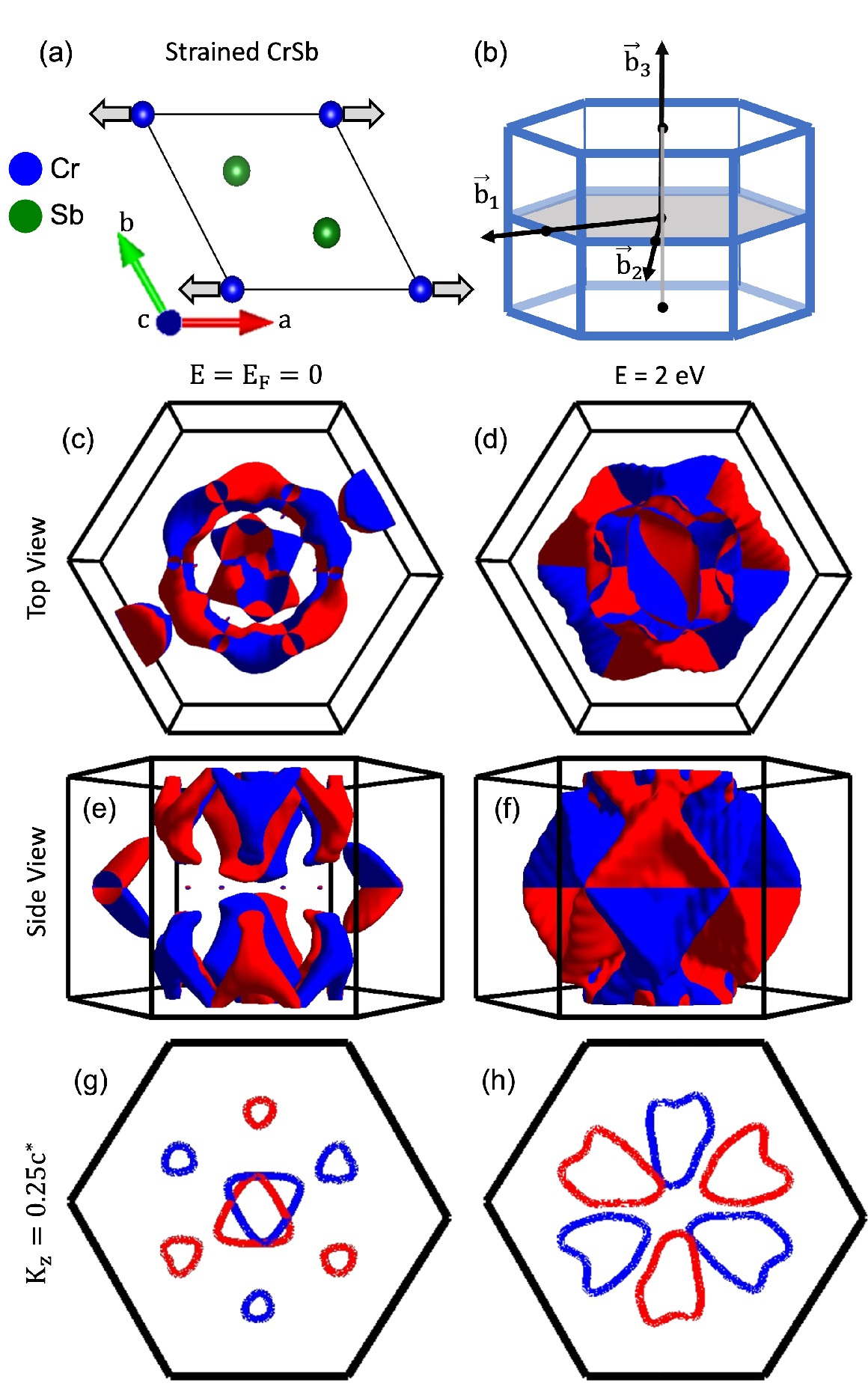}
    \caption{Altermagnetism in strained CrSb and the formation of FNCs in it. (a) A schematic illustration of the application of strain to pristine CrSb along $x$-axis. (b) The conversion of the diagonal nodal planes to a nodal line along $k_z$ axis due to reduction in symmetry. (c,d), (e,f), and (g,h) represent the top view, side view, and $k_x-k_y$ cross-sectional view of the constant energy surface, respectively, as explained in Fig. \ref{pristine_fermi_surface}.  While (c, e, g) represent the constant energy surface for $0$ eV, (d, f, h) represent the same for  $2$ eV. All the constant energy surface plots indicate the presence of FNCs in the strained CrSb.}
    \label{strained_CrSb}
\end{figure}

The formation of FNCs by restricting the symmetry to C$_{2z}$ and M$_z$ can be practically realized in the pristine CrSb itself. To achieve it, we have applied a uniaxial strain along $x$-axis which breaks the C$_{6z}$ symmetry. Here, we present the results for the case of 5\% strain ($a$ changes from 4.11 Å to 4.32 Å). To obtain the electronic and magnetic structures, the strain lattice parameter $a$ was kept fixed, while the system was relaxed to obtain new $b$ (= 4.13 Å) and $c$ (= 5.32 Å) as well as new atomic positions. Within this constraint, we find that the system retains the C$_{2z}$ and M$_z$ symmetry. The constant energy surface for E = E$_\text{F}$, which is shown in the upper panel of Fig.~\ref{strained_CrSb}. The figure demonstrates the deviation from the six-fold symmetry exhibited by the constant energy surface of the pristine CrSb (Fig.~\ref{pristine_fermi_surface}), and the formation of FNCs. From our analysis, we find that the increasing uniaxial strain further fragments the nodal curves. 

\subsection{Presence of FNCs in RbMnPO$_4$}

\begin{figure}[htbp]
    \centering
    \includegraphics[width=1.0\linewidth]{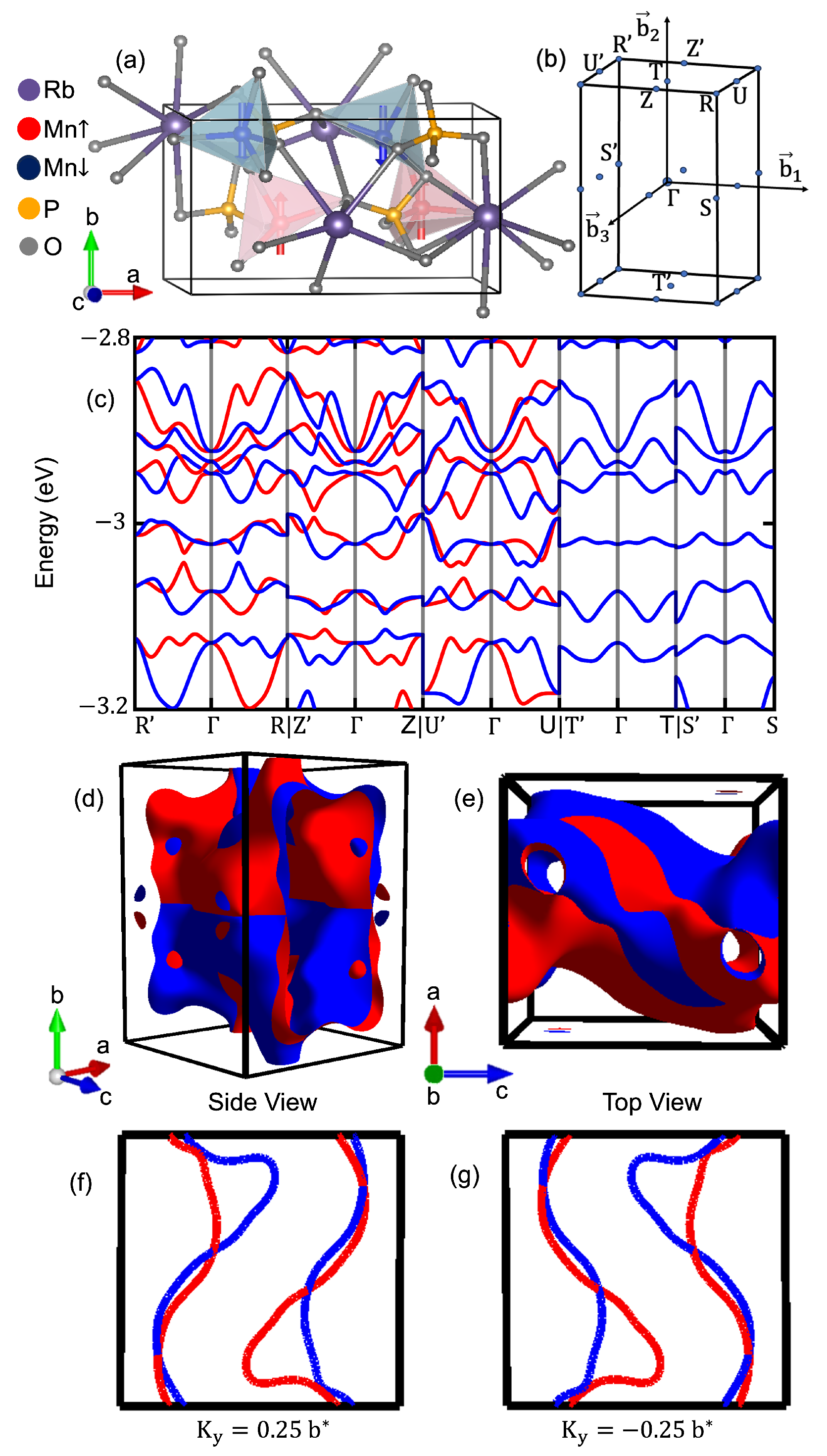}
    \caption{FNC driven altermagnetism in RbMnPO$_4$. (a) represents the crystal structure, (b) represents a schematic diagram of the first Brillouin zone with the high symmetry $k$-points, and (c) represents the altermagnetic band structure containing the basal nodal plane of RbMnPO$_4$. (d) and (e) show the side view and top view of the constant energy surface at $-3$ eV, respectively. (f) and (g) represent a $k_x-k_z$ cross-sectional view of the same for $k_y=0.25\, b*$ and $-0.25\, b*$, respectively.
    }
    \label{Experimental_Rb}
\end{figure}

To further validate the presence of FNCs, we scanned symmetry-lowered antiferromagnets where C$_{2x/2y/2z}$ is one of the symmetry elements. We identified insulating RbMnPO$_4$ as one such compound which is experimentally synthesized with a Néel temperature of 4K and an effective magnetic moment of 5.76 $\mu_B$ per Mn \cite{YAHIA200774}. The compound crystallizes in the SG P2$_1$ and has only one symmetry element C$_{2y}$t$_{(0 \frac{1}{2} 0)}$ in addition to identity. The SSG symbol for this system is, P$^{-1}$2$_1$$^{\infty m}$1. The experimental lattice parameters are  $a = 8.95$ Å, $b = 5.45$ Å, $c = 9.16$ Å and $\beta = 90.29^\circ$. The crystal structure (Fig.~\ref{Experimental_Rb} (a)) and the AFM band structure along certain altermagnetic k-paths shown in Fig.~\ref{Experimental_Rb} (c). The constant energy surface for E $= -3$ eV in Fig.~\ref{Experimental_Rb} (d) and (e) with respect to the Fermi level clearly demonstrates the presence of FNCs. Furthermore, the presence of basal nodal plane is attributed to the fact that in the NRL, C$_{2y}$ gives rise to the M$_y$ (Eqs.~\ref{Eq.7}-\ref{Eq.9}). The underlying mechanism is same as it is discussed in the context of MS-V. The lower panel of the Fig.~\ref{Experimental_Rb}, where the constant energy contours (E $= -3$ eV) are shown for k$_y = 0.25$ b$^*$ and k$_y = -0.25$b$^*$,  indicates that the spin-opposite sublattice bands intersect each other at specific points (nodal points) to ensure the validity of two fold rotation $C_2 || C_{2y}$. As k$_y$ varies continuously, these nodal points give rise to nodal curves. 

\section{Anomalous Hall Effect}

\begin{figure}[htbp]
    \centering
    \includegraphics[width=1\linewidth]{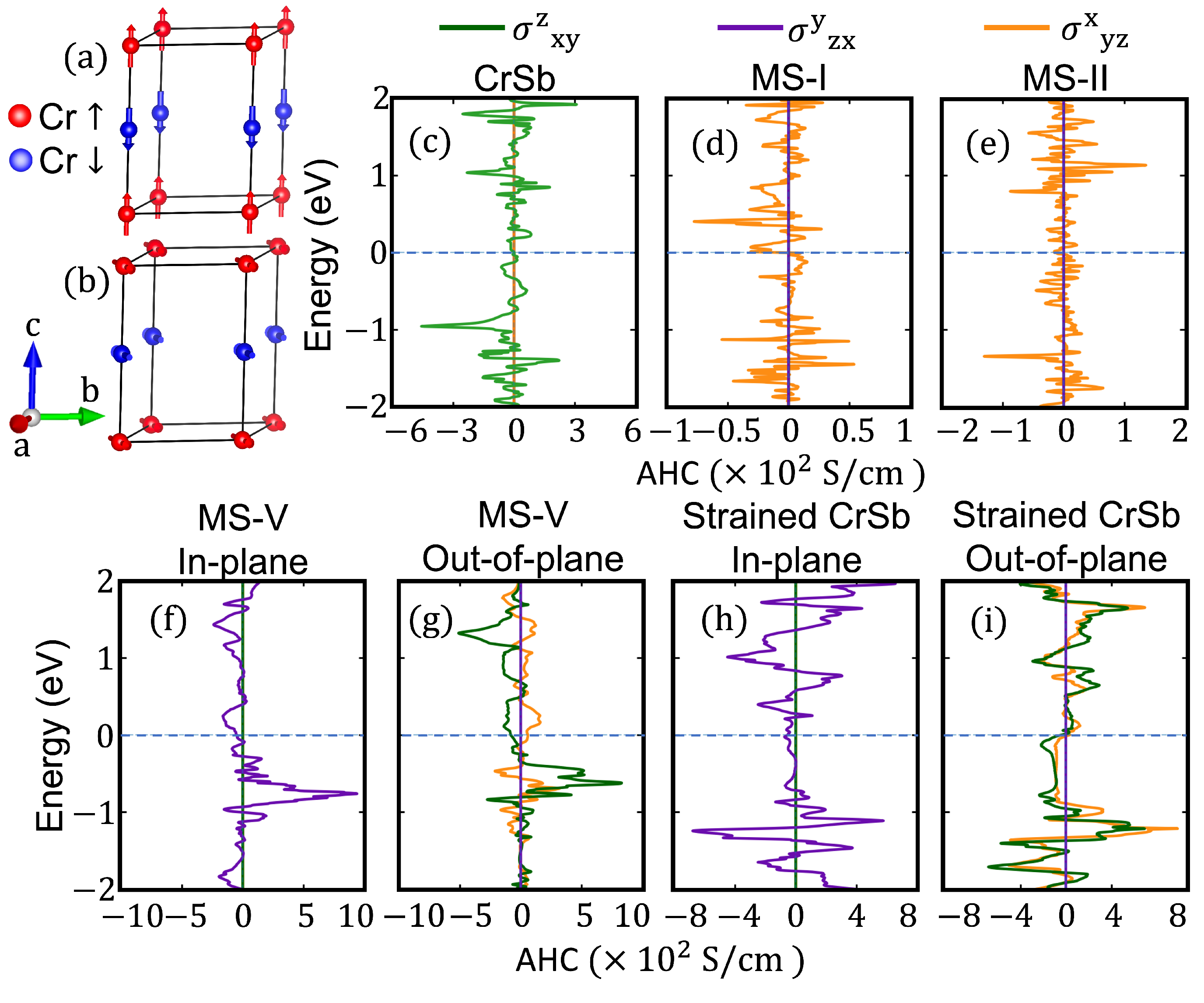}
    \caption{Anomalous Hall conductivity in pristine CrSb and various model structures. (a) and (b) schematically illustrate the orientation of spin quantization axes of the magnetic atoms in each configuration along out-of-plane ($z$-axis) and in-plane (with 30\textdegree{} angle from the crystallographic $a$-axis) directions, respectively. (c), (d), and (e) indicate the finite AHC of pristine CrSb, MS-I, and MS-II, respectively, when the Néel vector is aligned along the in-plane direction. (f) and (g) represent the finite AHC in MS-V for both in-plane and out-of-plane Néel vector orientation, respectively. (h) and (i) represent the same for the strained CrSb. While the model structures in the upper panel exhibit a finite AHC only in the in-plane Néel vector orientation, the model structures in the lower panel exhibit AHC both in in-plane and out-of-plane Néel vector orientations due to symmetry lowering.}
    \label{CrSb-dope_AHC}
\end{figure}

There is a wide range of reports suggesting that doping and other external stimuli can reorient the Néel vector and, thereby, modify the AHC~\cite{devaraj2025unlockingdopingeffectsaltermagnetism, fischer2026engineering, Yu2025}. Since these external factors in principle lower the symmetry of the system, the Cr$_m$Sb$_n$ MSs present an ideal platform to systematically establish the relationship between symmetry lowering and the corresponding changes in AHC, which can be estimated by using the Kubo formula \cite{PhysRevLett.49.405, doi:10.1143/JPSJ.12.570},

\begin{equation}
 \sigma_{\alpha \beta} = \frac{e^2}{\hbar} \sum_{n} \int_{BZ}^{} f(\epsilon_{n}(\textbf{k})) \Omega_{n}^{\gamma}(\textbf{k}) \frac{d\textbf{k}}{(2\pi)^3}.
 \label{sigma_eqn}
\end{equation}

Here, $e$, $\hbar$, $n$,  and $f(\epsilon_{n}(\textbf{k}))$ represent the electron charge, reduced Planck's constant, band index, and the Fermi-Dirac distribution function, respectively. The Berry curvature, $\Omega_{n}^{\nu}(\textbf{k})$ is expressed by the formula below,

\begin{equation}
    \Omega_n^\gamma (\boldsymbol{k}) = -2 \hbar^2 \  \text{Im} \sum_{n \neq n'} \frac{\langle \psi_{n'\boldsymbol{k}} | v_\alpha |\psi_{n\boldsymbol{k}} \rangle \langle \psi_{n\boldsymbol{k}} | v_\beta |\psi_{n'\boldsymbol{k}} \rangle}{(\epsilon_{n\boldsymbol{k}} - \epsilon_{n'\boldsymbol{k}})^2},          
    \label{berry_curv}
\end{equation}

Here, $v_\alpha(\beta) = (\hbar^{-1})\  \partial {\cal H}/\partial k_\alpha(\beta)$ is velocity operator,  $(\alpha, \beta, \gamma)$ are cyclic permutations of $(x, y, z)$, and $\epsilon_{n\boldsymbol{k}}$ is the energy eigenvalue of $n$-th band.

\begin{table*}[htbp]
\centering
\caption{
Symmetry analysis and anomalous Hall conductivity (AHC) tensor components for CrSb 
structures (Pristine, MS-I/MS-II, and MS-V) under different magnetic space groups (MSG) 
and Néel vector orientations. $E$ and $I$ denote identity and inversion, respectively; 
$C_{n,c}$ and $S_{n,c}$ represent $n$-fold rotation and rotoinversion about the $c$-axis; 
$M_n$ denotes mirror reflection normal to the $n$-axis; and $\tau$ indicates 
time-reversal combined with the symmetry operation.
}
\label{AHC-table}
\small

\begin{tabular*}{\textwidth}{@{\extracolsep{\fill}} l l l l c c c l c @{}}
\hline\hline
Structure & MSG & \shortstack{Néel Vector \\orientation} & Symmetries 
& $\sigma_{xy}^z$ & $\sigma_{zx}^y$ & $\sigma_{yz}^x$ 
& \shortstack{Possible AHC\\ Components} & \shortstack{Resulting AHC\\component}\\
\hline

% ---------------- PRISTINE ----------------
    Pristine CrSb & \shortstack{$P6_3'/m'm'c$\\194.268} & $c$-axis & $E, I$
& \checkmark & \checkmark & \checkmark & All & None \\
& & & $C_{3z}^{\pm}, S_{3z}^{\pm}$
& \checkmark & $\times$ & $\times$ & $\sigma_{xy}^z$ & \\
& & & $C_{2(210,120,1\bar{1}0)}$
& $\times$ & \checkmark & \checkmark & $\sigma_{zx}^y, \sigma_{yz}^x$ & \\
& & & $M_{210,120,1\bar{1}0}$
& $\times$ & \checkmark & \checkmark & $\sigma_{zx}^y, \sigma_{yz}^x$ & \\
& & & $\tau C_{6z}^{\pm}, \tau C_{2z}, \tau M_z$
& $\times$ & \checkmark & \checkmark & $\sigma_{zx}^y, \sigma_{yz}^x$ & \\
& & & $\tau C_{2x}, \tau M_x$
& \checkmark & \checkmark & $\times$ & $\sigma_{xy}^z, \sigma_{zx}^y$ & \\
& & & $\tau C_{2y}, \tau M_y$
& \checkmark & $\times$ & \checkmark & $\sigma_{xy}^z, \sigma_{yz}^x$ & \\
& & & $\tau C_{2(110)}, \tau M_{110}$
& \checkmark & \checkmark & \checkmark & All & \\
& & & $\tau S_{6z}^{\pm}$
& $\times$ & $\times$ & $\times$ & None & \\ \cline{2-9}

&\shortstack{$Cm'c'm$\\ 63.462} & $30^\circ$ w.r.t.\ $a$-axis & $E, I$
& \checkmark & \checkmark & \checkmark & All & $\sigma_{xy}^z$ \\
& & & $C_{2z}, M_z$
& \checkmark & $\times$ & $\times$ & $\sigma_{xy}^z$ & \\
& & & $\tau C_{2x}, \tau M_x$
& \checkmark & \checkmark & $\times$ & $\sigma_{xy}^z, \sigma_{zx}^y$ & \\
& & & $\tau C_{2y}, \tau M_y$
& \checkmark & $\times$ & \checkmark & $\sigma_{xy}^z, \sigma_{yz}^x$ & \\

\hline
% ---------------- MS-I / MS-II ----------------
MS-I/MS-II & \shortstack{$P\bar{6}^\prime m^\prime2$ \\(187.211)} & $c$-axis & $E, \tau M_{(110)}$
& \checkmark & \checkmark & \checkmark & All & None \\
& & & $C_{3z}^{\pm}$
& \checkmark & $\times$ & $\times$ & $\sigma_{xy}^z$ & \\
& & & $C_{2(210,120,1\bar{1}0)}$
& $\times$ & \checkmark & \checkmark & $\sigma_{zx}^y, \sigma_{yz}^x$ & \\
& & & $\tau S_{6z}^{\pm}$
& $\times$ & $\times$ & $\times$ & None & \\
& & & $\tau M_z$
& $\times$ & \checkmark & \checkmark & $\sigma_{zx}^y, \sigma_{yz}^x$ & \\
& & & $\tau M_x$
& \checkmark & \checkmark & $\times$ & $\sigma_{xy}^z, \sigma_{zx}^y$ & \\
& & & $\tau M_y$
& \checkmark & $\times$ & \checkmark & $\sigma_{xy}^z, \sigma_{yz}^x$ & \\ \cline{2-9}

& \shortstack{Amm$'$2$'$ \\(38.190)} & $30^\circ$ w.r.t.\ $a$-axis & $E$
& \checkmark & \checkmark & \checkmark & All & $\sigma_{yz}^x$ \\
& & & $M_x$
& $\times$ & $\times$ & \checkmark & $\sigma_{yz}^x$ & \\
& & & $\tau C_{2z}$
& $\times$ & \checkmark & \checkmark & $\sigma_{zx}^y, \sigma_{yz}^x$ & \\
& & & $\tau M_y$
& \checkmark & $\times$ & \checkmark & $\sigma_{xy}^z, \sigma_{yz}^x$ & \\

\hline
% ---------------- MS-V ----------------
MS-V/strained CrSb & \shortstack{P2$'$$_{1}$/m$'$\\(11.54)} & $c$-axis & $E, I$
& \checkmark & \checkmark & \checkmark & All 
& $\sigma_{xy}^z, \sigma_{yz}^x$ \\
& & & $\tau C_{2y}, \tau M_y$
& \checkmark & $\times$ & \checkmark 
& $\sigma_{xy}^z, \sigma_{yz}^x$ & \\ \cline{2-9}

& \shortstack{P2$_{1}$/m\\(11.50)} & $30^\circ$ w.r.t.\ $a$-axis & $E, I$
& \checkmark & \checkmark & \checkmark & All 
& $\sigma_{zx}^y$ \\
& & & $C_{2y}, M_y$
& $\times$ & \checkmark & $\times$ 
& $\sigma_{zx}^y$ & \\

\hline\hline
\end{tabular*}
\end{table*}

Depending on the orientation of the Néel vector, the MSG varies, and the Berry curvature distribution across the BZ is governed by the corresponding symmetries. These symmetry relations determine whether the AHC is allowed or forbidden in the system. We analyzed the symmetry operations of the pristine, strain, and MSs for Néel vector orientations along the $c$-axis and at 30\textdegree{} from the $a$-axis. The latter orientation of the Néel vector is chosen as it shows the maximum amount of AHC~\cite{Yu2025}. The presence or absence of the corresponding AHC components is summarized in Table~\ref{AHC-table}. The symmetry allowed AHC as a function of energy is shown in Fig.\ref{CrSb-dope_AHC}.

In pristine CrSb, experimental studies show that the Néel vector is oriented along the out-of-plane direction \cite{Reimers2024}. In this configuration, the system belongs to MSG ($P6_3'/m'm'c$) (No.~194.268). As mentioned in Table~\ref{AHC-table}, the symmetries present in the system enforce the cancellation of the Berry curvature throughout the Brillouin zone, resulting in a vanishing AHC tensor. When the Néel vector is aligned in-plane at 30\textdegree{}~ from $a$-axis, the magnetic space group symmetry lowers to $Cm'c'm$ (No.~63.462). In this reduced-symmetry configuration, the magnetic symmetries permit a non-zero $\sigma^{z}_{xy}$ component of AHC, whereas all other tensor components vanish by symmetry. Fig.~\ref{CrSb-dope_AHC} (c) shows AHC obtained in this case.

Next, we analyze the AHC for the considered model structures. Among the five of them, MS-I, MS-II, and MS-V are altermagnets and therefore have the potential to exhibit an AHC. In contrast, MS-III and MS-IV are antiferromagnetic phases in which the preservation of time-reversal symmetry enforces a complete cancellation of AHC for any Néel vector orientation. MS-I and MS-II belong to the MSG $P\bar{6}^\prime m^\prime2$ (No.~187.211) when the Néel vector is oriented along the out-of-plane direction, and to Amm$'$2$'$ (No.~38.190) when the Néel vector is along 30\textdegree{}~ from the $a$-axis. Similar to the pristine case, all components of AHC vanish in the out-of-plane Néel vector orientation. In contrast, for the in-plane Néel vector orientation, $\sigma^{x}_{yz}$ component is present. 

Among all model structures, MS-V exhibits the most distinct anomalous Hall behavior due to its reduced symmetry. Unlike the pristine system, MS-I, and MS-II, MS-V exhibits a finite AHC not only when the Néel vector is oriented in-plane, similar to the other systems, but also when it is oriented along the out-of-plane direction. The AHC corresponding to these orientations are present in Fig.~\ref{CrSb-dope_AHC} (f) and (g). While a single AHC component is present in the in-plane Néel vector orientation ($\sigma^{y}_{zx}$), two components of the AHC tensor ($\sigma^{x}_{yz}$ and $\sigma^{z}_{xy}$) are simultaneously present in the out-of-plane Néel vector orientation. 

The investigation on AHC in model structures provides a pathway to engineer AHC in CrSb. Our results show the symmetry lowering can be used as an effective route to vary AHC in CrSb. In particular, achieving symmetry of MS-V (SSG: P$^{-1}$2$_1$/$^{-1}$m$^{\infty m}$1, and MSG: P2$'$$_{1}$/m$'$) will enable finite AHC when the Néel vector is oriented along the out-of-plane direction. Applying a uniaxial strain on pristine CrSb along the $a/b$ crystallographic axis is one such method to achieve the symmetry of MS-V. To substantiate, in Fig.~\ref{CrSb-dope_AHC}-(h) and (i), we have shown the presence of non-zero AHC in the in-plane and out-of-plane Néel vector orientations for pristine CrSb under 5\% uniaxial strain along $a$-axis. Remarkably, the AHC value around the Fermi level in the strained CrSb is higher than that of pristine CrSb.

\section{Summary and Outlook}

To summarize, we carried out DFT calculations and symmetry analysis on CrSb, one of the widely investigated altermagnetic systems, to investigate the effect of symmetry lowering on the evolution of momentum dependent spin splitting and possible emergence of new quantum phenomena and transport. The symmetry lowering is achieved by designing hypothetical model structures through the insertion and removal of non-magnetic Sb at the void centers in these hexagonal systems. The model structures adhered to the well-established spin space group symmetry principles in predicting the presence or absence of altermagentic characteristics. Interestingly, when the symmetry is lowered to C$_{2z}$, the diagonal nodal planes vanish, and instead, fragmented nodal curves (FNCs) emerge, which had not been reported in the literature so far. Along an FNC, a pair of spin-opposite sublattice bands are degenerate, and elsewhere they are split with varying strength. Furthermore, the location of FNCs differ with each pair of such bands. Therefore, it opens up a wider scope to exploit the momentum space spin polarization (MSSP) for a tailored wave vector via chemical doping or field bias. We demonstrate that such a symmetry lowering in CrSb can be practically realized through uniaxial strain along the basal crystal axis. The presence of FNCs in general is further substantiated by calculating the magnetic structure of RbMnPO$_4$, which is a known experimentally synthesized antiferromagnet. The symmetry lowering also has a profound effect on the anomalous Hall conductivity (AHC). When the symmetry is restricted to C$_{2z}$, the strained CrSb exhibits finite anomalous Hall components when the Néel vector is directed both along out-of-plane and within the plane. Experimentally, it is known that for pristine CrSb, the Néel vector is out-of-plane and AHC is zero. The strain engineering, as presented in this work, is a viable tool to realize finite AHC, and thereby making this system promising in the field of spintronics.

\section*{Acknowledgement}
A.M. thanks MoE India for the PMRF fellowship.

\bibliography{ref.bib}
\end{document}